%% file: ICASSP25-AR_audio_inpainting-new_template.tex
\documentclass[conference]{IEEEtran}
\IEEEoverridecommandlockouts

\usepackage{amsmath,amssymb,amsfonts}
\usepackage{algorithmic}
\usepackage{graphicx}
\usepackage{textcomp}

\usepackage[textwidth=1.5cm, textsize=small]{todonotes}
\usepackage{xcolor}

\definecolor{linkred}{HTML}{a11616}
\definecolor{linkgreen}{HTML}{16a116}
\definecolor{linkblue}{HTML}{1616a1}
\usepackage{hyperref}
\hypersetup{
	colorlinks=true,
	linkcolor=linkred,
	filecolor=linkblue,
	urlcolor=linkblue,
	citecolor=linkgreen
}

\usepackage{booktabs}

\usepackage{tikz, pgfplots}
\pgfplotsset{compat=newest}
\usetikzlibrary{patterns}
\newlength\figurewidth
\graphicspath{{figures/}}
\pgfkeys{/pgf/number format/.cd, 1000 sep={\,}} 
\pgfplotsset{
	compat=newest,
	every axis/.append style={
		major grid style={line width=.5pt, draw=white!85!black},
		label style={font=\footnotesize},
		tick label style={font=\footnotesize},
		legend style={font=\footnotesize},
		title style={font=\small\bfseries}
	},
}
\usepackage{adjustbox}

\input{defs}

\begin{document}

\title{Tweaking autoregressive methods \\ for inpainting of gaps in audio signals
%
\thanks{The work was supported by the Czech Science Foundation (GA\v{C}R) Project No.\,23-07294S.}
}

\author{\IEEEauthorblockN{Ondřej Mokrý}
\IEEEauthorblockA{\textit{Department of Telecommunications} \\
	\textit{Brno University of Technology}\\
	Czech Republic}
\and
\IEEEauthorblockN{Pavel Rajmic}
\IEEEauthorblockA{\textit{Department of Telecommunications} \\
	\textit{Brno University of Technology}\\
	Czech Republic}
}

\maketitle

\begin{abstract}
A novel variant of the Janssen method for audio inpainting is presented
and compared to other popular audio inpainting methods based on autoregressive (AR) modeling.
Both conceptual differences and practical implications are discussed.
The experiments demonstrate the importance of the choice of the AR model estimator,
window/context length,
and model order.
The results show the superiority of the proposed gap-wise Janssen approach
using objective metrics,
which is confirmed by a listening test.	


\end{abstract}


\section{Introduction}
\label{sec:intro}

Audio inpainting is a challenging signal processing task, where missing parts of an audio signal have to be completed.
%
For a~human listener, the result should be as pleasant as possible and ideally 
free of artifacts.
Previously proposed audio inpainting solutions cover a~wide range of approaches, from autoregressive modeling
\cite{javevr86,Etter1996:Interpolation_AR,Kauppinen2002:reconstruction.method.long.portions.audio,Esquef2003:Interpolation.Long.Gaps.Warped.Burgs},
through optimization methods
\cite{Adler2012:Audio.inpainting,MokryZaviskaRajmicVesely2019:SPAIN,MokryRajmic2020:Inpainting.revisited,TanakaYatabeOikawa2024:PHAIN,Mokry2022:Audio.inpainting.NMF},
heuristics
\cite{lagrange2005long},
graph-based methods
\cite{Perraudin2018:Similarity.Graphs}
to deep learning
\cite{Moliner2024:Diffusion.Inpainting,Miotello2023:Deep.Prior.Inpainting.Harmonic.CNN,Marafioti2019:Context.encoder,Marafioti2021:GACELA}
and hybrid approaches \cite{Mezza2024:Hybrid.packet.loss.concealment}.

For signal gaps of up to ca 80 milliseconds, the iterative method of Janssen et al.\
\cite{javevr86} proposed in 1986 constantly ranks among the best, 
according to numerous studies
\cite{MokryZaviskaRajmicVesely2019:SPAIN,Mokry2022:Audio.inpainting.NMF,TaubockRajbamshiBalasz2021:SPAINMOD,Zaviska2023:Multiple.Hankel.matrix.inpainting}.
The extrapolation methods
\cite{Kauppinen2002:reconstruction.method.long.portions.audio,Esquef2003:Interpolation.Long.Gaps.Warped.Burgs,Kauppinen2002:Audio.signal.extrapolation,Roth2003:Frequency.Warped.Burg}
are non-iterative and utilize a~twofold extrapolation
(from left to right and
right to left)
while the two particular solutions are blended together using a~crossfading scheme.
Such an approach belongs to
the most popular, perhaps due to its simplicity and speed, and
is actually used in the Matlab function
\href{https://www.mathworks.com/help/signal/ref/fillgaps.html}{fillgaps}.
The patented method of Etter \cite{Etter1996:Interpolation_AR}
considers the just mentioned approach suboptimal and proposes
to 
aggregate
the two extrapolation directions 
in
a~single optimization criterion.

Besides slight variances in how to \emph{model the signal},
different algorithms for \emph{estimation of the coefficients} are also available
\cite{BrockwellDavis2006:Time.series},
which will be discussed in detail further on.

In this paper, we review the principle of autoregression-based methods,
point out the main differences between the particular popular approaches, and present
computational experiments on two audio datasets.
Most importantly,
we propose a~new variant of the Janssen algorithm \cite{javevr86} not present in the literature and examine its performance compared with known approaches.
This novel method does not rely on frame-wise signal processing,
in contrast to the original method.

This paper considers only gaps up to 80\,ms in length.
For larger gap sizes, autoregression usually starts to become in\-effi\-ci\-ent.
A~number of non-autoregressive methods cited above were designed to cope with larger gaps;
however, note that it is challenging to compare against such methods for at least two reasons:
first, they are typically trained on a~specific class of data, while AR modeling is data-independent;
second, these methods fill the gaps with material which may be pleasant to listen to,
but it does not have to align with the original audio, making it hard to objectively judge the reconstruction quality.

\section{Modeling audio as an autoregressive process}

%

A signal $\x = \transp{[x_1, \dots, x_N]}\in\RR^N$ is said to be modeled as an autoregressive (AR) process, if it satisfies%
\begin{equation}
	\sum_{i=1}^{p+1} a_{i} x_{n+1-i} = e_{n}, \quad n = 1,\dots, N+p,
	\label{eq:error.sum}
\end{equation}
where $\e = \transp{[e_1, \dots, e_{N+p}]}$ is a~realization of a~zero-mean white noise process, 
and $\a = \transp{[1,a_2, \dots, a_{p+1}]}$ are referred to as the AR coefficients
\cite[Def.\,3.1.2]{BrockwellDavis2006:Time.series}.
The order $p$ defines the range of indexes that determine the output in the current time instance.
As such, the frequency resolution increases with increasing $p$.
The above is closely related to the notion of convolution, and it is actually in accordance with the convention of the
\href{https://www.mathworks.com/help/signal/ref/lpc.html}{lpc} function in Matlab.

From the perspective of model fitting,
the vector $\e\in\RR^{N+p}$ is called the \emph{residual error}
\cite[Sec.\,8.2.2]{Zolzer2011:DAFX}.
Given the observed signal $\x$ and the order $p$, the coefficients of the AR model are usually estimated via the optimization problem
\begin{equation}
	\argmin_{\a} \frac{1}{2}\norm{\e(\a,\x)}^2\!,
	\label{eq:AR}
\end{equation}
where the error $\e(\a,\x)$, defined by its entries in \eqref{eq:error.sum}, is a~function of both the signal $\x$ and the coefficients $\a$.
Problem \eqref{eq:AR} can be effectively solved using
an estimate of the autocorrelation and the
Levinson--Durbin algorithm \cite{Durbin1960,Levinson1946}.
Such an approach is typically referred to as the LPC for its connection to the linear prediction coefficients.
\todo[disable]{PR: Následující věta mi přijde naráz bez kontextu, nechápu, co se jí chtělo říct...}

In contrast to the LPC, the Burg algorithm
\cite{Burg1975:PHD.Maximum.entropy.spectral}
for the estimation of the AR parameters involves an extra assumption that the \emph{same} parameters should model
both the signal $\x$ and its version flipped in time.
In effect, this results in another quadratic term extending \eqref{eq:AR}.
Kauppinen and Roth prefer the Burg algorithm for audio signal extrapolation \cite[Sec.\,4.2]{Kauppinen2002:Audio.signal.extrapolation}
since the underlying all-pole filter obtained this way is stable \cite[Sec.\,12.3.3]{Proakis1996:DSP}.
A~frequency-warped Burg algorithm has been proposed \cite{Roth2003:Frequency.Warped.Burg} that allows focusing on specified spectral bands;
however, the effect of warping is comparable to increasing the model order $p$  in the non-warped case \cite[Sec.\,5]{Roth2003:Frequency.Warped.Burg}.

Additional terms can be optionally appended to the objective \eqref{eq:AR} that \emph{regularizes} either the reconstructed signal or the AR coefficients, as proposed in \cite{Dufera2018:HOSpLP,Dufera2019:HOSpLP}.

\section{Audio inpainting using autoregression}


To formalize the problem of audio inpainting,
assume that the observed signal
$\x\in\RR^N$
\todo[disable]{PR: Zbytečné opakovat}
consists of
reliable
samples
identified by the set of indices $M\subset\{1,\dots,N\}$,
and vacant
samples at positions $\overline{M} = \{1,\dots,N\}\setminus M$.
The goal of inpainting is to estimate the missing samples at the positions $\overline{M}$.
In the so-called consistent case,
the samples at positions $M$ are meant to be preserved,
i.e., any candidate solution $\hat{\x}$ to the inpainting problem should satisfy $\hat{x}_n = x_n$ for all $n\in M$.

In this work,
the focus is on the practical scenario where the signal contains
gaps,
i.e., segments of consecutive lost samples,
surrounded by an intact context, as found in packet loss concealment, for instance
\cite{Mezza2024:Hybrid.packet.loss.concealment}.
For such a~scenario,
two
approaches based on AR modeling are applicable.

First,
the extrapolation-based method fits two independent sets of AR
parameters
for each gap,
one for the left-hand context and one for the right-hand context of the gap.
These coefficients are then used to extrapolate (predict) both contexts inside the gap,
and the forward- and backward-extrapolated signals are then cross-faded.
Numerous fading options are possible
(see, for instance, \cite[Sec.\,4.2]{Kauppinen2002:reconstruction.method.long.portions.audio});
we resort to the raised cosine function used in 
\cite{Etter1996:Interpolation_AR}.

Second,
the Janssen method operates independently on individual signal frames, typically obtained by windowing with overlaps.
In each frame,
the iterative Janssen method alternates between the estimation of the AR model for the frame (the current estimate of the missing samples being fixed)
and the missing samples in this frame~(with a~fixed estimate of the model parameters \cite{javevr86}.
As such,
a~problem similar to \eqref{eq:AR} is solved,
where both $\a$ and $\x$ are variables,
and $\x$ is constrained to stick to the reliable part of a~signal frame.
Using the overlap-add procedure, the individually processed frames are joined to form the output. 

Finally,
we propose a novel use of the Janssen method,
which,
instead of overlapping frames,
treats each gap in the signal separately (hence the name gap-wise Janssen).
Here, a~single AR model is 
simultaneously
fitted to \emph{both} left- and right-hand contexts of the gap.\!%
\footnote{%
	If, furthermore, the Burg algorithm is used to estimate the AR coefficients,
	it means that a single AR model is assumed not only for the whole gap context,
	but also for its flipped version.
}
Such a~way of using the context is similar to the extrapolation-based method,
but the AR coefficients (shared by both contexts) are estimated as in the Janssen method.
Note that neither the frame-wise or the proposed gap-wise Janssen approach is limited to the selected scenario where a~\emph{compact} gap is surrounded by reliable context.
This is, however, not the case of the extrapolation method, and therefore the experiments stick to that use case.

\section{Experiments \& Results}


To simulate degradation,
we consider gap lengths from 10 ms up to 80 ms, and create 10 gaps in each signal at pseudorandom locations.\!%
\footnote{Signals with fixed masks of the reliable samples are available in the repository
\url{https://github.com/ondrejmokry/TestSignals}.}
From AR-based methods,
we use all three aforementioned approaches:
The extrapolation-based method 
and the gap-wise Janssen method are applied with a~fixed context length of 4096 samples (approx.\ 93\,ms)
on each side of the gap (i.e., 8192 samples in total).
The frame-wise Janssen uses a~frame length of 4096 samples and two window shapes:
rectangular and Hann (see, e.g., \cite[Sec.\,V]{Harris1978}).
All methods are applied with the varying
model order $p$ and either the Burg or the LPC algorithm to fit the AR model.



The first measure of reconstruction quality is the signal-to-distortion ratio (SDR).
For the reference (i.e., undegraded) signal $\y$ and the reconstruction $\hat{\x}$, SDR in decibels is computed as
$\textup{SDR}(\y, \hat{\x}) = 10\log_{10}\frac{\norm{\y}^2}{\norm{\y-\hat{\x}}^2}$.
%
%
In our case, 
the SDR is only computed in the inpainted sections of the signal.
The perceived quality of the signal is assessed using PEMO-Q \cite{Huber:2006a},
an objective metric which predicts the subjective difference of $\y$ and $\hat{\x}$ in terms of the objective difference grade (ODG),
ranging from $-4$ (very annoying)
to~0 (imperceptible).  
An alternative choice, which is common in other audio processing fields, is PEAQ \cite{Thiede:2000a, Kabal2002:PEAQ}.
However,
this metric is not decisive
enough
in the case of gap inpainting \cite{MokryPhD2024}.

Our first dataset consists of 9 recordings of individual musical instruments%
\footnote{%
	We chose solo instruments since AR models are expected to perform well on them;
	a~sum of multiple AR processes generally may not be an AR process of order reasonably comparable with the individual signal orders.}\!,
taken from the EBU SQAM database~\cite{EBUSQAM}.
They are sampled at 44.1\,kHz and cut to a~length of around 7~seconds.
%
Note that further experiments were performed using a~longer window/context length of 8192 samples,
and also using a mid-scale dataset based upon the music IRMAS database
\cite{Bosch2012:IRMAS.paper, Bosch2014:IRMAS.dataset};
those results are presented later in Sec.~\ref{ssec:more.experiments}.


\begin{figure*}
	\hspace{0.08cm}%
	\adjustbox{scale=1.00}{\input{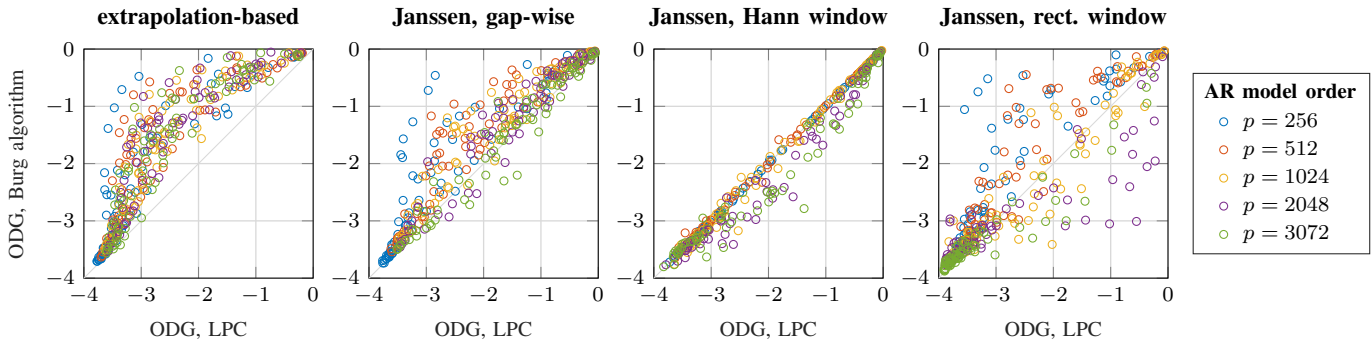}}%
	\vspace{-0.3cm}%
	\caption{%
		Comparison of the estimators in terms of PEMO-Q ODG for window/context length 4096 samples.
		Per each inpainting method,
		the scatter plot shows the individual results using LPC vs.\ the Burg algorithm to estimate the AR coefficients.
		The effect of the model order $p$ is analyzed separately in Fig.~\ref{fig:model_order}.
	}
	\label{fig:lpc_burg}
\end{figure*}

\begin{figure*}
	\hspace{0.07cm}%
	\adjustbox{scale=1.00}{\input{figures/model_order_pemoq.tex}}%
	\vspace{-0.3cm}%
	\caption{Comparison of the model order choices in terms of 
		PEMO-Q ODG for window/context length 4096 samples.
		Per each inpainting method,
		the plot shows averaged results using LPC (darker shade, dashed line) vs.\ the Burg algorithm (lighter shade, solid line) to estimate the AR coefficients.
	}
	\label{fig:model_order}
	\vspace{-0.25cm}
\end{figure*}
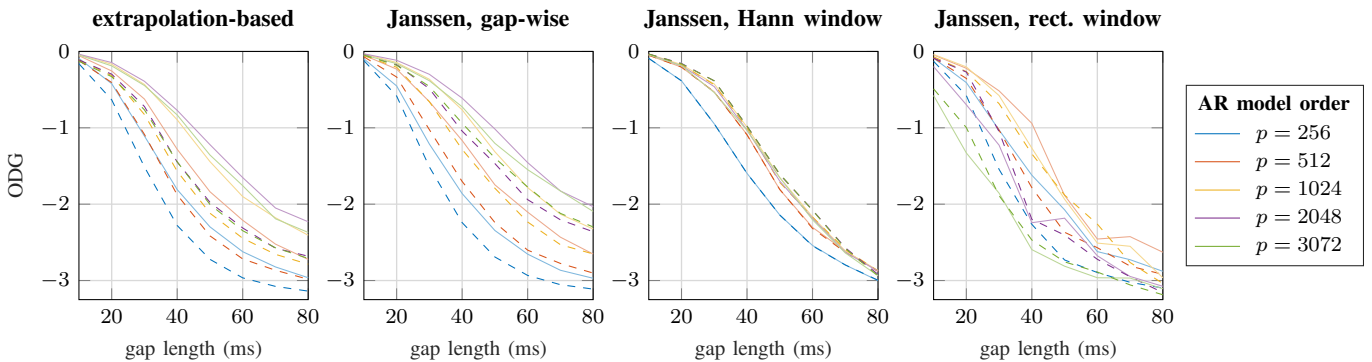

\subsection{Effect of the estimator}

First,
we evaluate the performance of the inpainting methods depending on the estimator of the AR model parameters.
Hence, for each test instance,
two versions of each inpainting method are run,
using either the LPC or the Burg algorithm.

The results are presented in Fig.~\ref{fig:lpc_burg}.
The overall distribution of the results in the scatter plots indicates that,
in terms of the ODG,
the Burg algorithm is clearly favorable in the case of the extrapolation-based inpainting.
The preferability of the Burg algorithm is also indicated in the gap-wise Janssen algorithm.
Notably, a~conclusion in the case of the frame-wise Janssen algorithm depends on the model order and the selected window.
With the Hann window,
inpainting results appear to depend on the chosen estimator only for large model orders
(LPC scores better in such cases).
With the rectangular window,
the dependence on the model order is clearly amplified.
Furthermore,
a~large portion of the results in this case implies
that the Burg algorithm is a~better choice if the model order is low, and vice versa.
The differences
have been verified as statistically significant using the Wilcoxon signed rank test \cite{Hollander1999}.

Note that in terms of SDR,
the differences are less pronounced compared to the ODG,
but the conclusions are analogous.
The complete results are available at the accompanying webpage.\!%
\footnote{\url{https://ondrejmokry.github.io/InpaintingAutoregressive/}}

%

\subsection{Effect of the model order}

The AR model order $p$ plays a crucial role in the modeling and also significantly affects the
results,
as demonstrated in Fig.\ \ref{fig:model_order}.
Note that Fig.~\ref{fig:model_order} plots the same data as Fig.~\ref{fig:lpc_burg},
but with the focus on the effect of the model order and the gap length.

Similarly to the study of preferences between LPC and the Burg algorithm,
the model order affects the results differently for different inpainting methods.
A clear scheme is observed in the case of the extrapolation-based and the gap-wise Janssen methods,
where increasing the model order up to $p=2048$ results in both a~higher SDR and a~higher ODG.
However,
the order $p=3072$ does not further increase the resulting quality. 
Note that the same observation holds for the experiment with window/context length 8192, including model order $p=4096$;
see the supplementary material online.
For the frame-wise Janssen,
the results are further affected by the chosen window shape:
the best results are achieved using
$p=512$ with
the rectangular window,
and $p=1024$ with the Hann window.

Furthermore,
Fig.~\ref{fig:model_order} reveals that the observed phenomena do not depend on the length of the gap.

\subsection{Comparison with other methods}

\begin{figure}
	\centering
	\adjustbox{width=0.70\linewidth}{\input{figures/with_spain_sdr.tex}}
	\\
	\adjustbox{width=0.70\linewidth}{\input{figures/with_spain_pemoq.tex}}
	\vspace{-1ex}%
	\caption{Comparison of the AR-based methods with SPAIN in terms of SDR (top) and ODG (bottom),
		averaged over all signals.
		In this experiment, all AR-based methods used the Burg algorithm to estimate the coefficients,
		using the best performing order $p$ according to the results reported in Fig.\ \ref{fig:model_order}.
	}
	\label{fig:with_spain}
\end{figure}
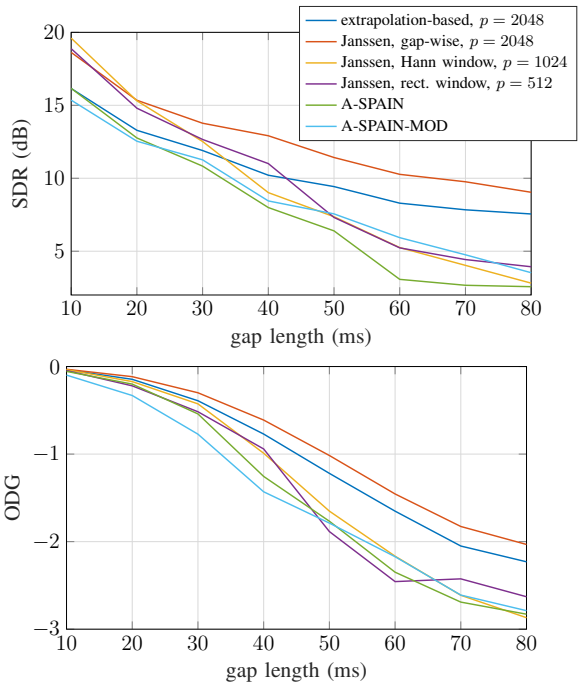

To provide a context for the results of AR-based methods,
selected variants are compared with the methods that belong to the state-of-the-art
in optimization-based audio inpainting,
namely A-SPAIN \cite{MokryZaviskaRajmicVesely2019:SPAIN}
and A-SPAIN-MOD \cite{TaubockRajbamshiBalasz2021:SPAINMOD},
applied with the same window length of 4096 samples.
The results in terms of SDR and ODG are presented in Fig.~\ref{fig:with_spain}.
The most significant observation is the dominance of the extrapolation-based and,
especially, the proposed gap-wise Janssen methods,
in particular in gaps larger than 50\,ms.


\begin{figure}
	\pgfplotsset{
		compat=newest,
		every axis/.append style={label style={font=\scriptsize},
			tick label style={font=\scriptsize},
			legend style={font=\scriptsize}}
	}
	\adjustbox{width=\linewidth}{\input{figures/boxplot.tex}}%
	\vspace{-2.5ex}%
	\caption{A boxplot showing the distribution of scores in the listening test.
	The proposed gap-wise Janssen method proves to be the best performing method,
	which is also confirmed statistically,
	since non-overlapping notches (filled areas) imply the difference of medians at the 5\% significance level.
	}%
	\label{fig:boxplot.subjective}%
\end{figure}
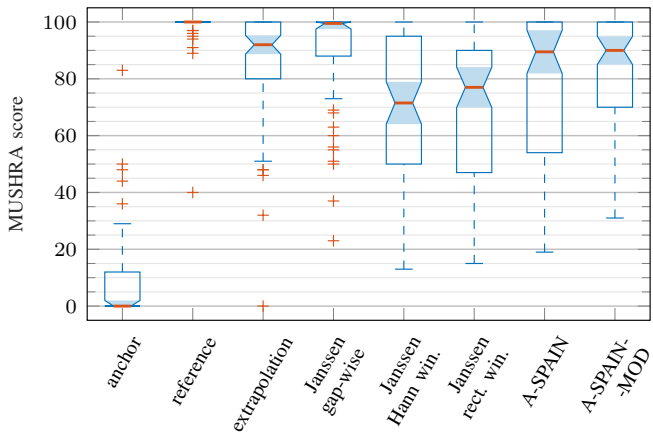

In addition to objective metrics, a~subjective listening test
was performed on a~subset of three audio excerpts
(the violin, piano, and clarinet)
with gaps of 20, 50 and 80 milliseconds in length, making 9~test signals altogether.
Such a~combination of signals, gap lengths and reconstruction methods yields
a~total length of 8.5 minutes of netto audio to evaluate.
We ran a~MUSHRA-type test
\cite{ITU-R2015:MUSHRA}, using the webMUSHRA environment \cite{Schoeffler2018:webMUSHRA}, 
in which the participants evaluated the quality of reconstructions.
The test was run in a~quiet music studio, using a~professional sound card and headphones.
The conditions were identical for all the participants.
Ten participants passed the pre- and postscreening selection.
We used the signal with gaps as the anchor.
%
%
The results, summarized in Fig.~\ref{fig:boxplot.subjective},
confirm the alignment of the subjective scores with the ODG metric,
presented in Fig.~\ref{fig:with_spain},
especially the ranking of the extrapolation-based and gap-wise Janssen methods.

\subsection{Testing on a mid-scale dataset, including increased window/context length}
\label{ssec:more.experiments}
%
An additional, mid-scale comparison was performed,
using 60 signals selected from the IRMAS database
\cite{Bosch2012:IRMAS.paper, Bosch2014:IRMAS.dataset}.
Our selection contains a~wide range of music characters and genres
but avoids recordings with highly pronounced drums.
Each final excerpt is 7~seconds long.
The test material is much more complex than in the solo-instrument case,
thus weakening the assumption about the autoregressive nature of signals,
which explains obtaining lower values of both SDR and ODG for all the methods considered.
However,
the results exhibit the same behavior as the small-scale experiment from Fig.~\ref{fig:with_spain};
in particular, the ranking of the methods was identical.

\begin{figure}
	\centering
	\adjustbox{width=0.70\linewidth}{\input{figures/with_spain_sdr_longer.tex}}%
	\vspace{-1ex}%
	\caption{Comparison of the AR-based methods with SPAIN in terms of SDR,
		using an increased window/context length of $8192$ samples on the solo-instrument dataset.
		All AR-based methods used the Burg algorithm to estimate the coefficients
		and the orders were selected as the best-performing for this window/context length.
		A-SPAIN-MOD is omitted for computational reasons (taking around 3.5 hours per signal).
	}
	\label{fig:with_spain_longer}
\end{figure}
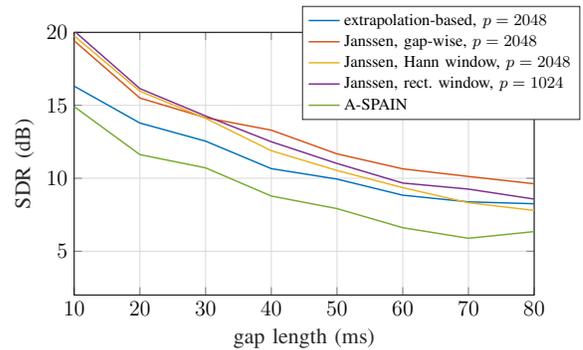

Finally,
all the experiments were repeated using a longer window/context length of 8192 samples.
This change is beneficial especially for the windowed methods, see Fig.\ \ref{fig:with_spain_longer};
however, the gap-wise Janssen remains superior on average.
For complete results, see the accompanying webpage.

\section{Conclusion}
A~gap-wise modification of the classic Janssen method was proposed and evaluated against popular audio inpainting methods based on autoregressive modeling.
The experiments demonstrated the importance of the choice of the AR model estimator
(i.e., choosing the LPC or the Burg algorithm)
and the model order.
The concluding tests, both objective and subjective, revealed that the gap-wise Janssen method (using the Burg algorithm)
is recommended as an autoregressive reference
for
inpainting of gaps up to 80\,ms;
this holds even in comparison with sparsity-based approaches.

If computational speed is an important criterion,
note that for all approaches, the computational load is proportional
both to the order of the AR model and to the gap length. 
Moreover, the Burg algorithm is more demanding compared to the LPC.
From the perspective of computational time%
\footnote{%
	The computations were performed on a desktop computer with
	Intel Core i7-10700K CPU at 3.80GHz
	with 32 GB RAM.}\!%
, the
extrapolation-based approach is clearly preferable
(elapsed times are up to around 0.15\,s per signal with $p=2048$,
while the gap-wise Janssen reaches up to 11.5\,s per signal with 
$p=1024$, and up to 16\,s with $p=2048$;
both using the Burg algorithm).

The Matlab codes for the methods discussed in the present paper
are publicly available.\!%
\footnote{\url{https://github.com/ondrejmokry/InpaintingAutoregressive}}

	

%

\clearpage

\input{hard_coded_bibliography.bbl}


%
%
%
%
%
%

\end{document}

%% file: defs.tex
\def\RR{\mathbb{R}}

\newcommand{\vect}[1]{\mathbf{#1}}
\def\x{\vect{x}}

\def\y{\vect{y}}

\def\a{\vect{a}}

\def\e{\vect{e}}

\newcommand{\norm}[1]{\|#1\|}

\newcommand{\transp}[1]{#1^\top}

\newcommand{\argmin}{\mathop{\operatorname{arg\,min}}}

\newcommand{\enlarge}{%
	\pgfplotsset{
		compat=newest,
		every axis plot/.append style={thick},
		every axis/.append style={label style={font=\large},
			tick label style={font=\large},
			legend style={font=\large}}
	}
}

%% file: figures/model_order_pemoq.tex
%
%
\definecolor{mycolor1}{rgb}{0.00000,0.44700,0.74100}%
\definecolor{mycolor2}{rgb}{0.85000,0.32500,0.09800}%
\definecolor{mycolor3}{rgb}{0.92900,0.69400,0.12500}%
\definecolor{mycolor4}{rgb}{0.49400,0.18400,0.55600}%
\definecolor{mycolor5}{rgb}{0.46600,0.67400,0.18800}%
\begin{tikzpicture}
	
	\begin{axis}[%
		width=1.20in,
		height=1.30in,
		at={(1.239in,0.443in)},
		scale only axis,
		xmin=10,
		xmax=80,
		xlabel style={text=white!15!black},
		xlabel={gap length (ms)},
		ymin=-3.25,
		ymax=0,
		ylabel style={text=white!15!black},
		ylabel={ODG},
		axis background/.style={fill=white},
		title={\raisebox{0pt}[0pt][0pt]{extrapolation-based}},
		xmajorgrids,
		ymajorgrids
		]
		\addplot [color=white!50!mycolor1, forget plot]
		table[row sep=crcr]{%
			10	-0.091624897612841\\
			20	-0.422286004041267\\
			30	-1.10788556813457\\
			40	-1.80898741759288\\
			50	-2.29701308247775\\
			60	-2.62244019356139\\
			70	-2.82179490306772\\
			80	-2.96331792413476\\
		};
		\addplot [color=mycolor1, dashed, forget plot]
		table[row sep=crcr]{%
			10	-0.161342178023256\\
			20	-0.63365619479051\\
			30	-1.51275112562836\\
			40	-2.27950583925387\\
			50	-2.72245724742632\\
			60	-2.96502396205091\\
			70	-3.07663616215856\\
			80	-3.13860167986245\\
		};
		\addplot [color=white!50!mycolor2, forget plot]
		table[row sep=crcr]{%
			10	-0.0553636449798979\\
			20	-0.255719108859286\\
			30	-0.62341929305567\\
			40	-1.28116812683006\\
			50	-1.83698801534932\\
			60	-2.21112775298384\\
			70	-2.52284123024733\\
			80	-2.72364071521607\\
		};
		\addplot [color=mycolor2, dashed, forget plot]
		table[row sep=crcr]{%
			10	-0.119339257342604\\
			20	-0.407216001478764\\
			30	-1.08642744027604\\
			40	-1.87834098563833\\
			50	-2.41500253039586\\
			60	-2.71692404274317\\
			70	-2.86536082673642\\
			80	-2.98340255801076\\
		};
		\addplot [color=white!50!mycolor3, forget plot]
		table[row sep=crcr]{%
			10	-0.0408920127440946\\
			20	-0.167560053249782\\
			30	-0.43266860734748\\
			40	-0.895451711707937\\
			50	-1.44229017121232\\
			60	-1.90075337584501\\
			70	-2.1778781567815\\
			80	-2.40696889919715\\
		};
		\addplot [color=mycolor3, dashed, forget plot]
		table[row sep=crcr]{%
			10	-0.111098064687732\\
			20	-0.315299372161037\\
			30	-0.825996306801661\\
			40	-1.56840515144943\\
			50	-2.11793894477388\\
			60	-2.44654984906446\\
			70	-2.65694831388803\\
			80	-2.78607358451876\\
		};
		\addplot [color=white!50!mycolor4, forget plot]
		table[row sep=crcr]{%
			10	-0.0361542424770202\\
			20	-0.145977626458787\\
			30	-0.390725684445011\\
			40	-0.771197153542244\\
			50	-1.22011183053611\\
			60	-1.6518515585394\\
			70	-2.04997239270736\\
			80	-2.23033913504493\\
		};
		\addplot [color=mycolor4, dashed, forget plot]
		table[row sep=crcr]{%
			10	-0.111426551544321\\
			20	-0.297559258294632\\
			30	-0.720944975787096\\
			40	-1.44820037937299\\
			50	-1.97530002266735\\
			60	-2.31163219537094\\
			70	-2.57638894627104\\
			80	-2.68165821255239\\
		};
		\addplot [color=white!50!mycolor5, forget plot]
		table[row sep=crcr]{%
			10	-0.0471864407998359\\
			20	-0.188901263965451\\
			30	-0.446471715063068\\
			40	-0.831576699886604\\
			50	-1.35586776092972\\
			60	-1.75789580665648\\
			70	-2.19141937910978\\
			80	-2.3700779859615\\
		};
		\addplot [color=mycolor5, dashed, forget plot]
		table[row sep=crcr]{%
			10	-0.12769963359745\\
			20	-0.330316959283319\\
			30	-0.764685332265169\\
			40	-1.44568049905963\\
			50	-2.0029160794953\\
			60	-2.34773733517583\\
			70	-2.57106410581048\\
			80	-2.710171330825\\
		};
	\end{axis}
	
	\begin{axis}[%
		width=1.20in,
		height=1.30in,
		at={(2.73in,0.443in)},
		scale only axis,
		xmin=10,
		xmax=80,
		xlabel style={text=white!15!black},
		xlabel={gap length (ms)},
		ymin=-3.25,
		ymax=0,
		axis background/.style={fill=white},
		title={\raisebox{0pt}[0pt][0pt]{Janssen, gap-wise}},
		xmajorgrids,
		ymajorgrids
		]
		\addplot [color=white!50!mycolor1, forget plot]
		table[row sep=crcr]{%
			10	-0.0964977335686655\\
			20	-0.45184435961522\\
			30	-1.21342831887209\\
			40	-1.86138887397171\\
			50	-2.34488242989337\\
			60	-2.65424378349706\\
			70	-2.86541040970231\\
			80	-2.9703407850727\\
		};
		\addplot [color=mycolor1, dashed, forget plot]
		table[row sep=crcr]{%
			10	-0.119693891186329\\
			20	-0.584056541778194\\
			30	-1.51254828045957\\
			40	-2.24162568491862\\
			50	-2.69167372208859\\
			60	-2.93304564013006\\
			70	-3.05422056725586\\
			80	-3.11226749934527\\
		};
		\addplot [color=white!50!mycolor2, forget plot]
		table[row sep=crcr]{%
			10	-0.0488499449355248\\
			20	-0.231625573673277\\
			30	-0.657409905366925\\
			40	-1.16606795589171\\
			50	-1.74662308086629\\
			60	-2.10393420428422\\
			70	-2.43197585256234\\
			80	-2.65309172203424\\
		};
		\addplot [color=mycolor2, dashed, forget plot]
		table[row sep=crcr]{%
			10	-0.0701185495921877\\
			20	-0.335217667410614\\
			30	-1.0246061825755\\
			40	-1.70291159867862\\
			50	-2.24033313769867\\
			60	-2.60208052345177\\
			70	-2.77584472434909\\
			80	-2.90226526918485\\
		};
		\addplot [color=white!50!mycolor3, forget plot]
		table[row sep=crcr]{%
			10	-0.0328926184071056\\
			20	-0.144789374704899\\
			30	-0.364632484433071\\
			40	-0.764636301025562\\
			50	-1.31358236849919\\
			60	-1.78284044509931\\
			70	-2.13127848817802\\
			80	-2.3249858481999\\
		};
		\addplot [color=mycolor3, dashed, forget plot]
		table[row sep=crcr]{%
			10	-0.0546607311946603\\
			20	-0.206971738446528\\
			30	-0.658891519654421\\
			40	-1.28343711896467\\
			50	-1.79145860803839\\
			60	-2.23326921560722\\
			70	-2.53046049263115\\
			80	-2.65188382352397\\
		};
		\addplot [color=white!50!mycolor4, forget plot]
		table[row sep=crcr]{%
			10	-0.0310579713470681\\
			20	-0.115663148090292\\
			30	-0.299561607363113\\
			40	-0.611299948812533\\
			50	-1.01605132215585\\
			60	-1.45545127708884\\
			70	-1.8268929455971\\
			80	-2.03219458741717\\
		};
		\addplot [color=mycolor4, dashed, forget plot]
		table[row sep=crcr]{%
			10	-0.0430837154092857\\
			20	-0.171565930621543\\
			30	-0.488403727043681\\
			40	-1.04352883060437\\
			50	-1.46831742743104\\
			60	-1.94091998823834\\
			70	-2.20481540571179\\
			80	-2.36101634895076\\
		};
		\addplot [color=white!50!mycolor5, forget plot]
		table[row sep=crcr]{%
			10	-0.0467670245509528\\
			20	-0.159538686308838\\
			30	-0.383622165670795\\
			40	-0.724655722246971\\
			50	-1.20331201953772\\
			60	-1.54810461987578\\
			70	-1.82696566673577\\
			80	-2.09530148343143\\
		};
		\addplot [color=mycolor5, dashed, forget plot]
		table[row sep=crcr]{%
			10	-0.0476146007602666\\
			20	-0.181630340380533\\
			30	-0.458253557203847\\
			40	-0.939367043752122\\
			50	-1.39510708355705\\
			60	-1.77546037504375\\
			70	-2.12039343224383\\
			80	-2.29787021700557\\
		};
	\end{axis}
	
	\begin{axis}[%
		width=1.20in,
		height=1.30in,
		at={(4.221in,0.443in)},
		scale only axis,
		xmin=10,
		xmax=80,
		xlabel style={text=white!15!black},
		xlabel={gap length (ms)},
		ymin=-3.25,
		ymax=0,
		axis background/.style={fill=white},
		title={\raisebox{0pt}[0pt][0pt]{Janssen, Hann window}},
		xmajorgrids,
		ymajorgrids,
		]
		\addplot [color=white!50!mycolor1]
		table[row sep=crcr]{%
			10	-0.0909177873864755\\
			20	-0.385014512374924\\
			30	-0.948388481396016\\
			40	-1.59636230955583\\
			50	-2.14195642850446\\
			60	-2.54047776713644\\
			70	-2.79596787825715\\
			80	-3.00067394709455\\
		};
		
		\addplot [color=mycolor1, dashed, forget plot]
		table[row sep=crcr]{%
			10	-0.0909221618945979\\
			20	-0.385193972173489\\
			30	-0.94834994940297\\
			40	-1.59688974167369\\
			50	-2.14252000605354\\
			60	-2.5458592420869\\
			70	-2.79573093227899\\
			80	-2.99628680288103\\
		};
		\addplot [color=white!50!mycolor2]
		table[row sep=crcr]{%
			10	-0.049849247473083\\
			20	-0.205140181578086\\
			30	-0.529739349112293\\
			40	-1.08934568401263\\
			50	-1.80916850782498\\
			60	-2.30415490677594\\
			70	-2.61376005929273\\
			80	-2.87595755412315\\
		};
		
		\addplot [color=mycolor2, dashed, forget plot]
		table[row sep=crcr]{%
			10	-0.0499843993040471\\
			20	-0.205274970081691\\
			30	-0.531433933881178\\
			40	-1.09262301822902\\
			50	-1.80682627231802\\
			60	-2.31796973199808\\
			70	-2.62953754624623\\
			80	-2.88004522145898\\
		};
		\addplot [color=white!50!mycolor3]
		table[row sep=crcr]{%
			10	-0.0353624252927701\\
			20	-0.17195667862217\\
			30	-0.42631650924447\\
			40	-0.990089533642467\\
			50	-1.65142619846824\\
			60	-2.16629078713027\\
			70	-2.61265479973694\\
			80	-2.8687951142124\\
		};
		
		\addplot [color=mycolor3, dashed, forget plot]
		table[row sep=crcr]{%
			10	-0.0362425173816665\\
			20	-0.174536131070864\\
			30	-0.436206524034441\\
			40	-1.00460845436208\\
			50	-1.7153834313001\\
			60	-2.16874126948142\\
			70	-2.59413829440439\\
			80	-2.87470586470069\\
		};
		\addplot [color=white!50!mycolor4]
		table[row sep=crcr]{%
			10	-0.0371173858892324\\
			20	-0.187715501014561\\
			30	-0.460828083200154\\
			40	-1.0291795179001\\
			50	-1.69787766092642\\
			60	-2.18465580887268\\
			70	-2.62828610577411\\
			80	-2.87717406472497\\
		};
		
		\addplot [color=mycolor4, dashed, forget plot]
		table[row sep=crcr]{%
			10	-0.0339176765056296\\
			20	-0.157298894936874\\
			30	-0.385324579077035\\
			40	-0.977007185839402\\
			50	-1.61225038973044\\
			60	-2.07709819456442\\
			70	-2.57040095670091\\
			80	-2.91988971126963\\
		};
		\addplot [color=white!50!mycolor5]
		table[row sep=crcr]{%
			10	-0.0441893582700554\\
			20	-0.186961949579952\\
			30	-0.538886994991888\\
			40	-1.03325503814482\\
			50	-1.65244256613998\\
			60	-2.20387113315135\\
			70	-2.63845945325361\\
			80	-2.93746256161126\\
		};
		
		\addplot [color=mycolor5, dashed, forget plot]
		table[row sep=crcr]{%
			10	-0.0344671851117238\\
			20	-0.162132753757856\\
			30	-0.384744767871868\\
			40	-0.978972363490243\\
			50	-1.60406338535243\\
			60	-2.07998512594796\\
			70	-2.57997554400299\\
			80	-2.93846950166631\\
		};
	\end{axis}
	
	\begin{axis}[%
		width=1.20in,
		height=1.30in,
		at={(5.712in,0.443in)},
		scale only axis,
		xmin=10,
		xmax=80,
		xlabel style={text=white!15!black},
		xlabel={gap length (ms)},
		ymin=-3.25,
		ymax=0,
		axis background/.style={fill=white},
		title={\raisebox{0pt}[0pt][0pt]{Janssen, rect.\ window}},
		xmajorgrids,
		ymajorgrids,
		legend style={legend cell align=left, align=left, draw=white!15!black, at={(1.1,0.5)}, anchor=west}
		]
		\addlegendimage{empty legend}
		\addlegendentry{\hspace{-5ex}\textbf{AR model order}}
		
		\addplot [color=white!50!mycolor1, forget plot]
		table[row sep=crcr]{%
			10	-0.0875386825090045\\
			20	-0.407951928513309\\
			30	-1.02854460990968\\
			40	-1.62045636746367\\
			50	-2.07406027864729\\
			60	-2.6211271844367\\
			70	-2.72738674617722\\
			80	-2.87924129454898\\
		};		
		\addplot [color=mycolor1, dashed, forget plot]
		table[row sep=crcr]{%
			10	-0.128963365143747\\
			20	-0.578303823259247\\
			30	-1.54465776983254\\
			40	-2.26433090887139\\
			50	-2.73070848700558\\
			60	-2.89525008726534\\
			70	-3.02331170414886\\
			80	-3.09789620887866\\
		};		
		\addplot [color=white!50!mycolor2, forget plot]
		table[row sep=crcr]{%
			10	-0.0459897278941861\\
			20	-0.220569489379194\\
			30	-0.516412656316135\\
			40	-0.941471216079882\\
			50	-1.88497940151254\\
			60	-2.4564845288054\\
			70	-2.42519769325549\\
			80	-2.62937470943985\\
		};		
		\addplot [color=mycolor2, dashed, forget plot]
		table[row sep=crcr]{%
			10	-0.0856304287051673\\
			20	-0.3532653096181\\
			30	-1.0318671360591\\
			40	-1.7834878935238\\
			50	-2.3681141634586\\
			60	-2.57021953106542\\
			70	-2.81092104058732\\
			80	-2.91970894081051\\
		};
		\addplot [color=white!50!mycolor3, forget plot]
		table[row sep=crcr]{%
			10	-0.0389173937053277\\
			20	-0.202122428338322\\
			30	-0.577848105636963\\
			40	-1.26550342338366\\
			50	-1.93514570339094\\
			60	-2.50950650043337\\
			70	-2.55157535372786\\
			80	-2.96418140259209\\
		};
		\addplot [color=mycolor3, dashed, forget plot]
		table[row sep=crcr]{%
			10	-0.0702909882228973\\
			20	-0.264027316729802\\
			30	-0.684169785158976\\
			40	-1.34127726687371\\
			50	-1.88111821088745\\
			60	-2.26119782543669\\
			70	-2.7638011985602\\
			80	-3.03865914644886\\
		};
		\addplot [color=white!50!mycolor4, forget plot]
		table[row sep=crcr]{%
			10	-0.198678466437445\\
			20	-0.698140644498399\\
			30	-1.2273466818436\\
			40	-2.24286320940538\\
			50	-2.18357740517224\\
			60	-2.67561608139705\\
			70	-2.94655988482717\\
			80	-3.07536249739185\\
		};
		\addplot [color=mycolor4, dashed, forget plot]
		table[row sep=crcr]{%
			10	-0.0871647422136499\\
			20	-0.269943123734258\\
			30	-1.05011994340091\\
			40	-2.197866626285\\
			50	-2.40509515516546\\
			60	-2.72193358301582\\
			70	-2.95030264234403\\
			80	-3.15740576417901\\
		};
		\addplot [color=white!50!mycolor5, forget plot]
		table[row sep=crcr]{%
			10	-0.576929867854218\\
			20	-1.32908971118937\\
			30	-1.85612569674361\\
			40	-2.59540573344698\\
			50	-2.81208536864532\\
			60	-2.96341845986609\\
			70	-2.96766204266031\\
			80	-3.10846580222985\\
		};
		\addplot [color=mycolor5, dashed, forget plot]
		table[row sep=crcr]{%
			10	-0.490107579775287\\
			20	-0.997248255032731\\
			30	-1.88865030931001\\
			40	-2.46898127079469\\
			50	-2.75551856181784\\
			60	-2.88835537276775\\
			70	-3.05755459942502\\
			80	-3.18644049823809\\
		};
	
		\addplot [color=mycolor1]
		table[row sep=crcr]{%
			0 0\\
			1 0\\
		};
		\addlegendentry{\hspace{0.5em}$p = 256$}
		
		\addplot [color=mycolor2]
		table[row sep=crcr]{%
			0 0\\
			1 0\\
		};
		\addlegendentry{\hspace{0.5em}$p = 512$}
		
		\addplot [color=mycolor3]
		table[row sep=crcr]{%
			0 0\\
			1 0\\
		};
		\addlegendentry{\hspace{0.5em}$p = 1024$}
		
		\addplot [color=mycolor4]
		table[row sep=crcr]{%
			0 0\\
			1 0\\
		};
		\addlegendentry{\hspace{0.5em}$p = 2048$}
		
		\addplot [color=mycolor5]
		table[row sep=crcr]{%
			0 0\\
			1 0\\
		};
		\addlegendentry{\hspace{0.5em}$p = 3072$}
	\end{axis}
	
\end{tikzpicture}%

%% file: figures/with_spain_sdr.tex
%
%
\definecolor{mycolor1}{rgb}{0.00000,0.44700,0.74100}%
\definecolor{mycolor2}{rgb}{0.85000,0.32500,0.09800}%
\definecolor{mycolor3}{rgb}{0.92900,0.69400,0.12500}%
\definecolor{mycolor4}{rgb}{0.49400,0.18400,0.55600}%
\definecolor{mycolor5}{rgb}{0.46600,0.67400,0.18800}%
\definecolor{mycolor6}{rgb}{0.30100,0.74500,0.93300}%
\enlarge
\begin{tikzpicture}[trim axis left, trim axis right]
	
	\begin{axis}[%
		width=3.5in,
		height=2.0in,
		at={(0.758in,0.481in)},
		scale only axis,
		xmin=10,
		xmax=80,
		xlabel style={text=white!15!black},
		xlabel={gap length (ms)},
		ymin=2,
		ymax=20,
		ylabel style={text=white!15!black},
		ylabel={SDR (dB)},
		axis background/.style={fill=white},
		xmajorgrids,
		ymajorgrids,
		legend style={at={(1.10,1.10)}, anchor=north east, legend cell align=left, align=left, draw=white!15!black, font=\small}
		]
		\addplot [color=mycolor1]
		table[row sep=crcr]{%
			10	16.1512418874303\\
			20	13.2864987910056\\
			30	11.9000268478401\\
			40	10.2082906132531\\
			50	9.42930248953038\\
			60	8.29283103648075\\
			70	7.83850468988678\\
			80	7.55247493059489\\
		};
		\addlegendentry{extrapolation-based, $p = 2048$}
		
		\addplot [color=mycolor2]
		table[row sep=crcr]{%
			10	18.6259135872454\\
			20	15.3457042779823\\
			30	13.7716787910352\\
			40	12.9128228990526\\
			50	11.4221370877055\\
			60	10.2685157361508\\
			70	9.75943177362145\\
			80	9.03881200155933\\
		};
		\addlegendentry{Janssen, gap-wise, $p = 2048$}
		
		\addplot [color=mycolor3]
		table[row sep=crcr]{%
			10	19.6036946891058\\
			20	15.3074625758231\\
			30	12.5190794691485\\
			40	9.01615819218254\\
			50	7.36432496305197\\
			60	5.24890028333693\\
			70	4.03074544425506\\
			80	2.81211199536543\\
		};
		\addlegendentry{Janssen, Hann window, $p = 1024$}
		
		\addplot [color=mycolor4]
		table[row sep=crcr]{%
			10	18.8804568785158\\
			20	14.8006274084239\\
			30	12.6594993911984\\
			40	11.0056385293903\\
			50	7.31414007265942\\
			60	5.22950933717408\\
			70	4.42809967891603\\
			80	3.93545329253067\\
		};
		\addlegendentry{Janssen, rect.\ window, $p = 512$}
		
		\addplot [color=mycolor5]
		table[row sep=crcr]{%
			10	16.1616145307965\\
			20	12.7620775978667\\
			30	10.8316150101077\\
			40	7.99293840595634\\
			50	6.39418461906147\\
			60	3.0713319102271\\
			70	2.66568413703025\\
			80	2.56831411762501\\
		};
		\addlegendentry{A-SPAIN}
		
		\addplot [color=mycolor6]
		table[row sep=crcr]{%
			10	15.354265723374\\
			20	12.538895159837\\
			30	11.2617454229969\\
			40	8.44647269574381\\
			50	7.55886226515842\\
			60	5.92950461657826\\
			70	4.75469897329019\\
			80	3.52686733075153\\
		};
		\addlegendentry{A-SPAIN-MOD}
		
	\end{axis}
	
\end{tikzpicture}%

%% file: figures/with_spain_pemoq.tex
%
%
\definecolor{mycolor1}{rgb}{0.00000,0.44700,0.74100}%
\definecolor{mycolor2}{rgb}{0.85000,0.32500,0.09800}%
\definecolor{mycolor3}{rgb}{0.92900,0.69400,0.12500}%
\definecolor{mycolor4}{rgb}{0.49400,0.18400,0.55600}%
\definecolor{mycolor5}{rgb}{0.46600,0.67400,0.18800}%
\definecolor{mycolor6}{rgb}{0.30100,0.74500,0.93300}%
\enlarge
\begin{tikzpicture}[trim axis left, trim axis right]
	
	\begin{axis}[%
		width=3.5in,
		height=2.0in,
		at={(0.758in,0.481in)},
		scale only axis,
		xmin=10,
		xmax=80,
		xlabel style={text=white!15!black},
		xlabel={gap length (ms)},
		ymin=-3,
		ymax=0,
		ylabel style={text=white!15!black},
		ylabel={ODG},
		axis background/.style={fill=white},
		xmajorgrids,
		ymajorgrids,
		]
		\addplot [color=mycolor1]
		table[row sep=crcr]{%
			10	-0.0361542424770202\\
			20	-0.145977626458787\\
			30	-0.390725684445011\\
			40	-0.771197153542244\\
			50	-1.22011183053611\\
			60	-1.6518515585394\\
			70	-2.04997239270736\\
			80	-2.23033913504493\\
		};
		
		\addplot [color=mycolor2]
		table[row sep=crcr]{%
			10	-0.0310579713470681\\
			20	-0.115663148090292\\
			30	-0.299561607363113\\
			40	-0.611299948812533\\
			50	-1.01605132215585\\
			60	-1.45545127708884\\
			70	-1.8268929455971\\
			80	-2.03219458741717\\
		};
		
		\addplot [color=mycolor3]
		table[row sep=crcr]{%
			10	-0.0353624252927701\\
			20	-0.17195667862217\\
			30	-0.42631650924447\\
			40	-0.990089533642467\\
			50	-1.65142619846824\\
			60	-2.16629078713027\\
			70	-2.61265479973694\\
			80	-2.8687951142124\\
		};
		
		\addplot [color=mycolor4]
		table[row sep=crcr]{%
			10	-0.0459897278941861\\
			20	-0.220569489379194\\
			30	-0.516412656316135\\
			40	-0.941471216079882\\
			50	-1.88497940151254\\
			60	-2.4564845288054\\
			70	-2.42519769325549\\
			80	-2.62937470943985\\
		};
		
		\addplot [color=mycolor5]
		table[row sep=crcr]{%
			10	-0.0556506158616167\\
			20	-0.200682237145153\\
			30	-0.5412033818748\\
			40	-1.25514473787089\\
			50	-1.76678641952239\\
			60	-2.34903899445193\\
			70	-2.69062061608635\\
			80	-2.82875841697913\\
		};
		
		\addplot [color=mycolor6]
		table[row sep=crcr]{%
			10	-0.0979117144940085\\
			20	-0.33091914009669\\
			30	-0.772592112927194\\
			40	-1.43057298074778\\
			50	-1.78816550587453\\
			60	-2.17213020106928\\
			70	-2.61035578917115\\
			80	-2.78865155975933\\
		};
		
	\end{axis}
\end{tikzpicture}%

%% file: figures/boxplot.tex
%
%
\definecolor{mycolor1}{rgb}{0.75000,0.86175,0.93525}%
\definecolor{mycolor2}{rgb}{0.00000,0.44700,0.74100}%
\definecolor{mycolor3}{rgb}{0.85000,0.32500,0.09800}%
\begin{tikzpicture}

\begin{axis}[%
width=2.7in,
height=1.5in,
at={(0.758in,0.605in)},
scale only axis,
unbounded coords=jump,
xmin=0.5,
xmax=8.5,
xtick={1,2,3,4,5,6,7,8},
xticklabel style={align=center},
xticklabels={{anchor},{reference},{extrapolation},{Janssen \\ gap-wise},{Janssen \\ Hann win.},{Janssen \\ rect.\ win.},{A-SPAIN},{A-SPAIN-\\-MOD}},
xticklabel style={rotate=60, align=right},
ymin=-5,
ymax=105,
ytick={0,20,...,100},
ylabel style={text=white!15!black},
ylabel={MUSHRA score},
axis background/.style={fill=white},
ymajorgrids,
yminorgrids,
major grid style={white!75!black},
minor grid style={white!90!black},
minor y tick num=3,
]

\addplot[area legend, draw=none, fill=mycolor1, forget plot]
table[row sep=crcr] {%
x	y\\
0.875	0\\
0.75	1.98591037058574\\
1.25	1.98591037058574\\
1.125	0\\
1.25	0\\
0.75	0\\
0.875	0\\
}--cycle;

\addplot[area legend, draw=none, fill=mycolor1, forget plot]
table[row sep=crcr] {%
x	y\\
7.875	90\\
7.75	94.9647759264644\\
8.25	94.9647759264644\\
8.125	90\\
8.25	85.0352240735356\\
7.75	85.0352240735356\\
7.875	90\\
}--cycle;

\addplot[area legend, draw=none, fill=mycolor1, forget plot]
table[row sep=crcr] {%
x	y\\
2.875	92\\
2.75	95.3098506176429\\
3.25	95.3098506176429\\
3.125	92\\
3.25	88.6901493823571\\
2.75	88.6901493823571\\
2.875	92\\
}--cycle;

\addplot[area legend, draw=none, fill=mycolor1, forget plot]
table[row sep=crcr] {%
x	y\\
3.875	99.5\\
3.75	100\\
4.25	100\\
4.125	99.5\\
4.25	97.5140896294143\\
3.75	97.5140896294143\\
3.875	99.5\\
}--cycle;

\addplot[area legend, draw=none, fill=mycolor1, forget plot]
table[row sep=crcr] {%
x	y\\
4.875	71.5\\
4.75	78.9471638896965\\
5.25	78.9471638896965\\
5.125	71.5\\
5.25	64.0528361103035\\
4.75	64.0528361103035\\
4.875	71.5\\
}--cycle;

\addplot[area legend, draw=none, fill=mycolor1, forget plot]
table[row sep=crcr] {%
x	y\\
5.875	77\\
5.75	84.1161788279322\\
6.25	84.1161788279322\\
6.125	77\\
6.25	69.8838211720678\\
5.75	69.8838211720678\\
5.875	77\\
}--cycle;

\addplot[area legend, draw=none, fill=mycolor1, forget plot]
table[row sep=crcr] {%
x	y\\
6.875	89.5\\
6.75	97.1126564205787\\
7.25	97.1126564205787\\
7.125	89.5\\
7.25	81.8873435794213\\
6.75	81.8873435794213\\
6.875	89.5\\
}--cycle;
\addplot [color=mycolor2, dashed, forget plot]
  table[row sep=crcr]{%
1	12\\
1	29\\
};
\addplot [color=mycolor2, dashed, forget plot]
  table[row sep=crcr]{%
2	100\\
2	100\\
};
\addplot [color=mycolor2, dashed, forget plot]
  table[row sep=crcr]{%
3	100\\
3	100\\
};
\addplot [color=mycolor2, dashed, forget plot]
  table[row sep=crcr]{%
4	100\\
4	100\\
};
\addplot [color=mycolor2, dashed, forget plot]
  table[row sep=crcr]{%
5	95\\
5	100\\
};
\addplot [color=mycolor2, dashed, forget plot]
  table[row sep=crcr]{%
6	90\\
6	100\\
};
\addplot [color=mycolor2, dashed, forget plot]
  table[row sep=crcr]{%
7	100\\
7	100\\
};
\addplot [color=mycolor2, dashed, forget plot]
  table[row sep=crcr]{%
8	100\\
8	100\\
};
\addplot [color=mycolor2, dashed, forget plot]
  table[row sep=crcr]{%
1	0\\
1	0\\
};
\addplot [color=mycolor2, dashed, forget plot]
  table[row sep=crcr]{%
2	100\\
2	100\\
};
\addplot [color=mycolor2, dashed, forget plot]
  table[row sep=crcr]{%
3	51\\
3	80\\
};
\addplot [color=mycolor2, dashed, forget plot]
  table[row sep=crcr]{%
4	73\\
4	88\\
};
\addplot [color=mycolor2, dashed, forget plot]
  table[row sep=crcr]{%
5	13\\
5	50\\
};
\addplot [color=mycolor2, dashed, forget plot]
  table[row sep=crcr]{%
6	15\\
6	47\\
};
\addplot [color=mycolor2, dashed, forget plot]
  table[row sep=crcr]{%
7	19\\
7	54\\
};
\addplot [color=mycolor2, dashed, forget plot]
  table[row sep=crcr]{%
8	31\\
8	70\\
};
\addplot [color=mycolor2, forget plot]
  table[row sep=crcr]{%
0.875	29\\
1.125	29\\
};
\addplot [color=mycolor2, forget plot]
  table[row sep=crcr]{%
1.875	100\\
2.125	100\\
};
\addplot [color=mycolor2, forget plot]
  table[row sep=crcr]{%
2.875	100\\
3.125	100\\
};
\addplot [color=mycolor2, forget plot]
  table[row sep=crcr]{%
3.875	100\\
4.125	100\\
};
\addplot [color=mycolor2, forget plot]
  table[row sep=crcr]{%
4.875	100\\
5.125	100\\
};
\addplot [color=mycolor2, forget plot]
  table[row sep=crcr]{%
5.875	100\\
6.125	100\\
};
\addplot [color=mycolor2, forget plot]
  table[row sep=crcr]{%
6.875	100\\
7.125	100\\
};
\addplot [color=mycolor2, forget plot]
  table[row sep=crcr]{%
7.875	100\\
8.125	100\\
};
\addplot [color=mycolor2, forget plot]
  table[row sep=crcr]{%
0.875	0\\
1.125	0\\
};
\addplot [color=mycolor2, forget plot]
  table[row sep=crcr]{%
1.875	100\\
2.125	100\\
};
\addplot [color=mycolor2, forget plot]
  table[row sep=crcr]{%
2.875	51\\
3.125	51\\
};
\addplot [color=mycolor2, forget plot]
  table[row sep=crcr]{%
3.875	73\\
4.125	73\\
};
\addplot [color=mycolor2, forget plot]
  table[row sep=crcr]{%
4.875	13\\
5.125	13\\
};
\addplot [color=mycolor2, forget plot]
  table[row sep=crcr]{%
5.875	15\\
6.125	15\\
};
\addplot [color=mycolor2, forget plot]
  table[row sep=crcr]{%
6.875	19\\
7.125	19\\
};
\addplot [color=mycolor2, forget plot]
  table[row sep=crcr]{%
7.875	31\\
8.125	31\\
};
\addplot [color=mycolor2, forget plot]
  table[row sep=crcr]{%
0.875	0\\
0.75	1.98591037058574\\
0.75	12\\
1.25	12\\
1.25	1.98591037058574\\
1.125	0\\
1.25	0\\
1.25	0\\
0.75	0\\
0.75	0\\
0.875	0\\
};
\addplot [color=mycolor2, forget plot]
  table[row sep=crcr]{%
1.875	100\\
1.75	100\\
1.75	100\\
2.25	100\\
2.25	100\\
2.125	100\\
2.25	100\\
2.25	100\\
1.75	100\\
1.75	100\\
1.875	100\\
};
\addplot [color=mycolor2, forget plot]
  table[row sep=crcr]{%
2.875	92\\
2.75	95.3098506176429\\
2.75	100\\
3.25	100\\
3.25	95.3098506176429\\
3.125	92\\
3.25	88.6901493823571\\
3.25	80\\
2.75	80\\
2.75	88.6901493823571\\
2.875	92\\
};
\addplot [color=mycolor2, forget plot]
  table[row sep=crcr]{%
3.875	99.5\\
3.75	100\\
3.75	100\\
4.25	100\\
4.25	100\\
4.125	99.5\\
4.25	97.5140896294143\\
4.25	88\\
3.75	88\\
3.75	97.5140896294143\\
3.875	99.5\\
};
\addplot [color=mycolor2, forget plot]
  table[row sep=crcr]{%
4.875	71.5\\
4.75	78.9471638896965\\
4.75	95\\
5.25	95\\
5.25	78.9471638896965\\
5.125	71.5\\
5.25	64.0528361103035\\
5.25	50\\
4.75	50\\
4.75	64.0528361103035\\
4.875	71.5\\
};
\addplot [color=mycolor2, forget plot]
  table[row sep=crcr]{%
5.875	77\\
5.75	84.1161788279322\\
5.75	90\\
6.25	90\\
6.25	84.1161788279322\\
6.125	77\\
6.25	69.8838211720678\\
6.25	47\\
5.75	47\\
5.75	69.8838211720678\\
5.875	77\\
};
\addplot [color=mycolor2, forget plot]
  table[row sep=crcr]{%
6.875	89.5\\
6.75	97.1126564205787\\
6.75	100\\
7.25	100\\
7.25	97.1126564205787\\
7.125	89.5\\
7.25	81.8873435794213\\
7.25	54\\
6.75	54\\
6.75	81.8873435794213\\
6.875	89.5\\
};
\addplot [color=mycolor2, forget plot]
  table[row sep=crcr]{%
7.875	90\\
7.75	94.9647759264644\\
7.75	100\\
8.25	100\\
8.25	94.9647759264644\\
8.125	90\\
8.25	85.0352240735356\\
8.25	70\\
7.75	70\\
7.75	85.0352240735356\\
7.875	90\\
};
\addplot [color=mycolor3, line width=1.0pt, forget plot]
  table[row sep=crcr]{%
0.875	0\\
1.125	0\\
};
\addplot [color=mycolor3, line width=1.0pt, forget plot]
  table[row sep=crcr]{%
1.875	100\\
2.125	100\\
};
\addplot [color=mycolor3, line width=1.0pt, forget plot]
  table[row sep=crcr]{%
2.875	92\\
3.125	92\\
};
\addplot [color=mycolor3, line width=1.0pt, forget plot]
  table[row sep=crcr]{%
3.875	99.5\\
4.125	99.5\\
};
\addplot [color=mycolor3, line width=1.0pt, forget plot]
  table[row sep=crcr]{%
4.875	71.5\\
5.125	71.5\\
};
\addplot [color=mycolor3, line width=1.0pt, forget plot]
  table[row sep=crcr]{%
5.875	77\\
6.125	77\\
};
\addplot [color=mycolor3, line width=1.0pt, forget plot]
  table[row sep=crcr]{%
6.875	89.5\\
7.125	89.5\\
};
\addplot [color=mycolor3, line width=1.0pt, forget plot]
  table[row sep=crcr]{%
7.875	90\\
8.125	90\\
};
\addplot [color=black, only marks, mark=+, mark options={solid, draw=mycolor3}, forget plot]
  table[row sep=crcr]{%
1	36\\
1	44\\
1	48\\
1	50\\
1	83\\
};
\addplot [color=black, only marks, mark=+, mark options={solid, draw=mycolor3}, forget plot]
  table[row sep=crcr]{%
2	40\\
2	89\\
2	91\\
2	94\\
2	95\\
2	96\\
2	97\\
2	97\\
};
\addplot [color=black, only marks, mark=+, mark options={solid, draw=mycolor3}, forget plot]
  table[row sep=crcr]{%
3	0\\
3	32\\
3	46\\
3	48\\
3	48\\
};
\addplot [color=black, only marks, mark=+, mark options={solid, draw=mycolor3}, forget plot]
  table[row sep=crcr]{%
4	23\\
4	37\\
4	50\\
4	51\\
4	55\\
4	56\\
4	60\\
4	63\\
4	68\\
4	69\\
};
\addplot [color=black, only marks, mark=+, mark options={solid, draw=mycolor3}, forget plot]
  table[row sep=crcr]{%
nan	nan\\
};
\addplot [color=black, only marks, mark=+, mark options={solid, draw=mycolor3}, forget plot]
  table[row sep=crcr]{%
nan	nan\\
};
\addplot [color=black, only marks, mark=+, mark options={solid, draw=mycolor3}, forget plot]
  table[row sep=crcr]{%
nan	nan\\
};
\addplot [color=black, only marks, mark=+, mark options={solid, draw=mycolor3}, forget plot]
  table[row sep=crcr]{%
nan	nan\\
};
\end{axis}

\end{tikzpicture}%

%% file: figures/with_spain_sdr_longer.tex
%
%
\definecolor{mycolor1}{rgb}{0.00000,0.44700,0.74100}%
\definecolor{mycolor2}{rgb}{0.85000,0.32500,0.09800}%
\definecolor{mycolor3}{rgb}{0.92900,0.69400,0.12500}%
\definecolor{mycolor4}{rgb}{0.49400,0.18400,0.55600}%
\definecolor{mycolor5}{rgb}{0.46600,0.67400,0.18800}%
\definecolor{mycolor6}{rgb}{0.30100,0.74500,0.93300}%
\enlarge
\begin{tikzpicture}[trim axis left, trim axis right]
	
	\begin{axis}[%
		width=3.5in,
		height=2.0in,
		at={(0.758in,0.481in)},
		scale only axis,
		xmin=10,
		xmax=80,
		xlabel style={text=white!15!black},
		xlabel={gap length (ms)},
		ymin=2,
		ymax=20,
		ylabel style={text=white!15!black},
		ylabel={SDR (dB)},
		axis background/.style={fill=white},
		xmajorgrids,
		ymajorgrids,
		legend style={at={(1.10,1.10)}, anchor=north east, legend cell align=left, align=left, draw=white!15!black, font=\small}
		]
		\addplot [color=mycolor1]
		table[row sep=crcr]{%
			10	16.3123157310574\\
			20	13.786396248349\\
			30	12.5430746599711\\
			40	10.6596204558892\\
			50	9.94360570756123\\
			60	8.83972444023777\\
			70	8.38225638728932\\
			80	8.25426481845599\\
		};
		\addlegendentry{extrapolation-based, $p = 2048$}
		
		\addplot [color=mycolor2]
		table[row sep=crcr]{%
			10	19.4216962785405\\
			20	15.4983352057854\\
			30	14.152291994162\\
			40	13.2971791639402\\
			50	11.6707816317072\\
			60	10.6458695146405\\
			70	10.1228847904562\\
			80	9.62311985073428\\
		};
		\addlegendentry{Janssen, gap-wise, $p = 2048$}
		
		\addplot [color=mycolor3]
		table[row sep=crcr]{%
			10	19.7542903115849\\
			20	15.9672963629197\\
			30	14.1110557668054\\
			40	11.8874175926441\\
			50	10.5320700477529\\
			60	9.3578241869686\\
			70	8.32284339899406\\
			80	7.79686385571889\\
		};
		\addlegendentry{Janssen, Hann window, $p = 2048$}
		
		\addplot [color=mycolor4]
		table[row sep=crcr]{%
			10	20.092538322466\\
			20	16.1448878555291\\
			30	14.268950894318\\
			40	12.5032617782452\\
			50	11.0145740652054\\
			60	9.67207860632689\\
			70	9.25709671671856\\
			80	8.57439526005988\\
		};
		\addlegendentry{Janssen, rect.\ window, $p = 1024$}
		
		\addplot [color=mycolor5]
		table[row sep=crcr]{%
			10	14.9109730665785\\
			20	11.621211043725\\
			30	10.713588540451\\
			40	8.78120886289056\\
			50	7.92082703842147\\
			60	6.61198795922672\\
			70	5.88682589398958\\
			80	6.34550983094978\\
		};
		\addlegendentry{A-SPAIN}
		
	\end{axis}
	
\end{tikzpicture}%

%% file: hard_coded_bibliography.bbl
\newcommand{\noopsort}[1]{} \newcommand{\printfirst}[2]{#1}
  \newcommand{\singleletter}[1]{#1} \newcommand{\switchargs}[2]{#2#1}